\documentclass[twoside]{article}
\usepackage{Proc_NTSE_18}
\addtocounter{page}{309}
\begin{document}
\thispagestyle{plain}
%Please use the command \publref{myfilename} to print the reference to your proceedings contribution 
%at the bottom of the page where myfilename should be replaced by the name of your LaTeX file
%(e.g., use the command \publref{Johns} if the LaTeX file of your contribution called Johns.tex):
\publref{MazurI}

\begin{center}
{\Large \bf \strut \boldmath
%Insert the title of your contribution here
Elastic $n{-}{^6\rm He}$ Scattering and $^7$He Resonant States\\ 
in the No-Core Shell Model\label{igor}
\strut}\\
\vspace{10mm}
{\large \bf\strut 
%Insert the authors here. Use upper indexes a, b, c, etc., to bind authors with their addresses
% as shown below.
I. A. Mazur$^{a,b}$, A. M. Shirokov$^{b,c,d}$,  I. J. Shin$^e$, A. I. Mazur$^b$, Y.~Kim$^e$, P. Maris$^d$ and J. P. Vary$^d$\strut}
\end{center}

\noindent{\small $^a$\it Center for Extreme Nuclear Matters, Korea University, Seoul 02841, Republic of Korea} \\
{\small $^b$\it Department of Physics, Pacific National University, Khabarovsk 680035, Russia} \\
{\small $^c$\it Skobeltsyn\hspace{-1pt} Institute\hspace{-1pt} of\hspace{-1pt} Nuclear\hspace{-1pt}
 Physics,\hspace{-1pt} Lomonosov\hspace{-1pt} Moscow\hspace{-1pt} State\hspace{-1pt} 
 University,\hspace{-1pt} Moscow\hspace{-1pt} 119991, 
\phantom{$^c$}Russia} \\
{\small $^d$\it Department of Physics and Astronomy, Iowa State University, Ames, IA 50011-3160, USA}
{\small $^e$\it Rare Isotope Science Project, Institute for Basic Science, Daejeon 305-811, Republic of\\ \phantom{$^c$}Korea}

%The next command defines running titles:
\markboth{
%Put here the list of authors that will be displayed in running titles:
I. A. Mazur, A. M. Shirokov, I. J. Shin, A. I. Mazur, Y.~Kim, P. Maris, J. P. Vary}
{%Put here the short title of your contribution that will be displayed in running titles:
Elastic $n{-}{^6\!He}$  scattering and $^7$\!He resonant states in the no-core shell model} 

\begin{abstract}
We present  results of calculations of  $n{-}{^6\rm He}$  elastic 
scattering phase shifts %partial waves
 and resonances in $^7$He. % obtained in SS-HORSE method. C
 The calculations  utilize the SS-HORSE method combined with %based on 
 {\it ab initio} no-core shell model calculations of the %results for 
 $^7$He and $^6$He nuclei with %the 
 Daejeon16 and the JISP16 $NN$ interactions.
\\[\baselineskip] 
\end{abstract}

\section{Introduction}

A modern trend of nuclear theory is a development of methods
{%
for describing} nuclear states in the continuum, resonances in particular, % (especially the resonances),
 as well as {%\clr
  the boundaries} of nuclear stability 
and nuclei beyond the drip lines. Obviously, %a design of 
{\it ab initio} (``%from 
first-principles'') approaches in this field {%\clr
 are} of %a
primary
%a great 
importance. %{\it Ab initio} means that the 
The only input for {\it ab initio} 
theoretical studies is the nucleon-nucleon ($NN$) and, if needed, three-nucleon 
($3N$) interactions.

Currently there {%\clr
 are} a number of %are 
reliable methods for {\it ab initio} description of 
nuclear bound states (see, e.\:g., the review~\cite{Leidemann}){. Prominent methods include}
the Green function's Monte Carlo~\cite{GFMC}, 
the no-core shell model (NCSM)~\cite{NCSM}, the coupled cluster 
method~\cite{CCM}, etc. The NCSM calculations are utilized %in the NCSM are used 
in this paper. 
The NCSM is a modern 
version of the nuclear shell model which does not introduce an inert core  
and {includes} the degrees of freedom of all nucleons of a given nucleus. 
The multi-particle wave function is expanded in a % the
 series of basis
many-body oscillator functions (Slater determinants) which include all many-body 
oscillator 
states with total excitation quanta less or equal to some given {value defined in terms of} 
$N_{\max}$. This
makes it possible to separate completely the center-of-mass motion. The 
number of basis states increases very rapidly with number of nucleons $A$ 
{and with}~$N_{\max}${. 
The} achievement of a reasonable accuracy of the NCSM 
calculations {is primarily} limited by the memory of available modern 
leadership-class supercomputers. {Currently, NCSM applications are obtained for} nuclei with the number of nucleons of
about~20. {As  $A$ increases, due to computational limits restricting basis space sizes,  there is a greater need for extrapolations to estimate converged results.}

However, the NCSM cannot be directly applied to the description of resonant 
states. Energies of resonant states are positive with respect to some
{threshold so that one needs}
to consider decay modes. Special methods taking into account the 
continuum {are therefore} needed for the description of resonances.

There are well-developed methods for {\it ab initio} description of 
continuum spectrum states based on Faddeev and Faddeev--Yakubovsky 
equations that are successfully applied in nuclear physics for systems with 
$A\leq5$ nucleons (see, e.\:g., the review~\cite{Leidemann}
and Ref.~\cite{Rimantas}). 
{A very} important breakthrough in developing {\it ab initio} theory of nuclear 
reactions in systems with 
total number of nucleons $A > 4$ was achieved by combining the NCSM and
%There are NCSM extensions based on 
the resonating 
group method {to built the so-called NCSM with continuum (NCSMC) approach}% {methods}
~\cite{Navratil} which has been applied to description of several 
nuclear systems with up to 11% nucleons
~\cite{Navratil_Be11} {and very recently up to 12~nucleons~\cite{talkPetr}}. Nuclear 
resonances can be considered also in the no-core Gamow shell model 
(GSM)~\cite{NCGSM}. 
However, these methods bring forth additional challenges for a numerical 
realization and the respective
calculations become % are 
very {demanding}. % for numeric calculations.

Recently we proposed the SS-HORSE 
method~\cite{SSHORSE, SSHORSE_PEPAN,Blkh,Blokh, SSHORSE-K}, which 
generalizes the NCSM to the continuum spectrum states. The SS-HORSE 
{allows one to calculate the} single-channel $S$-matrix and resonances by 
a simple analysis of NCSM eigenenergy behavior as a function of parameters 
of the many-body oscillator basis. The SS-HORSE extension of the NCSM was 
successfully applied
{to the} calculation of the neutron--$\alpha$ and proton--$\alpha$  scattering and 
resonant states in
the $^{5}$He and $^{5}$Li nuclei in Refs.~\cite{SSHORSE, SSHORSE-K}; a 
generalization of this
approach {to} the case of the democratic decay provided a description {of a} resonance
in the system of four neutrons (tetraneutron)~\cite{tetran}.

A brief review of the SS-HORSE {method is} presented in 
Section~\ref{Sec:sshorse}. Results for a single-channel neutron scattering by the
$^6$He nucleus and resonances in the $^7$He nucleus are presented in 
Section~\ref{Sec:n-6He}.

\section{SS-HORSE method\label{Sec:sshorse}}

Consider a channel of neutron scattering by a nucleus with $A$ nucleons.
%For the
The phase shift calculations  within the SS-HORSE approach start from the 
calculation of the set of
%it is needed to use 
the NCSM eigenenergies 
$E_{i}^{A+1}$ with some set of the NCSM basis parameters~$N^{i}_{\max}$ 
and~$\hbar\Omega^{i}$ for the whole %full 
(${A+1}$)-particle system, as well as of the ground state 
energies~$E_{i}^A$ of  the target nucleus with the same~$\hbar\Omega^{i}$ and the 
excitation 
quanta~$N^{i}_{\max}$ or~$N^{i}_{\max}-1$ depending of the parity of the states 
of interest of the (${A+1}$)-particle system. 
The respective relative motion energy is the difference
\begin{equation}
	\label{EnergyRelative}
	E_{i}=E_{i}^{A+1}-E_{i}^{A}. 
\end{equation}

The phase shifts~$\delta_\ell(E_{i})$ at the eigenenergies~$E_{i}$
 in the %of 
partial wave with the orbital momentum $\ell$ in the case of neutral particle 
scattering
%long-range Coulomb interaction absence 
are calculated as~\cite{SSHORSE,SSHORSE_PEPAN,Blkh} 
%with the help of 
%formula
\begin{equation}
	\label{PhaseSSHORSE}
	\tan \delta_\ell(E_{i})=-\frac{S_{{\mathbb N}^{i}+2, \ell}(E_{i})}
	                          {C_{{\mathbb N}^{i}+2, \ell}(E_{i})}.
\end{equation}
Here $S_{n,\ell}(E)$ and $C_{n,\ell}(E)$ are the regular and irregular 
oscillator solutions for the free motion, their analytical expressions can be 
found in 
Refs.~\cite{yamani,zaytmf,Bang};
%are known~\cite{Bang};
the oscillator quanta of the relative motion
\begin{equation}
	\label{QuantaRelative}
	{\mathbb N}^{i}=N^{i}_{\max}+N^{A+1}_{\min}-N^{A}_{\min}, 
\end{equation}
where $N^{i}_{\max}$ is the excitation quanta in the $(A+1)$-particle 
system in the current calculation, $N^{A+1}_{\min}$ and $N^{A}_{\min}$ are the 
minimal total oscillator quanta
consistent with the Pauli principle in the 
$(A+1)$- and $A$-particle systems, respectively. The energies~{$E_{i}$} depend, % from 
of course, on the NCSM 
basis parameters, $N^{i}_{\max}$ and $\hbar\Omega^{i}$. Therefore by varying 
these %this 
parameters (note, $\hbar\Omega$ appears in the definition of the functions 
$S_{n,\ell}$ 
and~$C_{n,\ell}$) we can calculate the %receive 
phase shifts in some energy interval. 
Next we perform the phase shift parameterization which makes it possible to 
calculate the $S$-matrix 
and its poles including those associated with the resonant states in the 
$(A+1)$-body system.
%We 
%get a information about $S$-matrix poles (include resonance poles) through 
%parameterization of obtained phase shifts.

%\mbox{}
%
%{\clr{}HERE}
%
%\mbox{}

The phase shifts can be parameterized using %with the help of 
the effective range function,
\begin{equation}
	\label{EffRadius}
	K(E)=\left(\sqrt{2\mu E}/\hbar\right)^{2\ell+1}\cot\delta_\ell(E),
\end{equation}
where $\mu$ is the reduced mass of scattered particles. 
The function~\eqref{EffRadius} has % have
 good analytical properties and may be expanded in
%expressed to 
Taylor series of energy $E$ (the so-called effective range expansion),
\begin{equation}
K(E)=-\frac{1}{a_{\ell}}+\frac{\mu r_{\ell}}{\hbar^{2}}E+cE^{2}+...\,,
\label{effrengeexp}
\end{equation}
where $a_{\ell}$ is the scattering length and~$r_{\ell}$ is the effective {range}. 
The 
expansion~\eqref{effrengeexp} works well at low energies, however in a larger 
energy interval, in 
particular, in the region of a resonance, it may be inadequate since  the phase 
shift may may take the values
of 0, $\pm\pi$, $\pm2\pi$,\:..., when the effective range function~$K(E)$, 
according to Eq.~\eqref{EffRadius},
tends to infinity. Therefore we %use 
express the effective range function as a
Pad\'e approximant,
%\begin{equation}
%	\label{Pade}
%	K(E)=\frac{w^{(n)}_0+w^{(n)}_1E+w^{(n)}_2E^2+...}%\cdots}
%	{1+w^{(d)}_1E+w^{(d)}_2E^2+...}. %\cdots}
%\end{equation}
\begin{equation}
	\label{Pade}
	K(E)=\frac{-1+w^{(n)}_1E+w^{(n)}_2E^2+...}%\cdots}
	{a_{\ell}+w^{(d)}_1E+w^{(d)}_2E^2+...}. %\cdots}
\end{equation}
Clearly, at low energies the Pad\'e approximant~\eqref{Pade} unambiguously 
transforms into
the effective range expansion~\eqref{effrengeexp}.
%Note, %since 
%the phase shift may may take the values %has 
%0, $\pm\pi$, $\pm2\pi$,\:... when the effective range function~$K(E)$, according to Eq.~\eqref{EffRadius},
%tends to infinity.
%% values in energy interval 
%%under considering.

With any set of parameters~$w^{(n)}_1$, $w^{(n)}_2,\:...\,$, $a_{\ell}$,
$w^{(d)}_1$, $w^{(d)}_2,\:...$ parametrizing the effective range function~$K(E)$ 
we can easily calculate
the phase shifts~$\delta_\ell(E)$ in the energy interval of interest and 
calculate the 
energies~$E^{th}_{i}$ using 
Eq.~\eqref{PhaseSSHORSE} for any combination of the NCSM
parameters~$N^{i}_{\max}$ and~$\hbar\Omega^{i}$. These energies~$E^{th}_{i}$ are 
compared
with the set of energies~$E_{i}$ obtained in the NCSM calculations; 
%The algorithm for finding 
the optimal values of $w^{(n)}_1$, $w^{(n)}_2,\:...\,$, $a_{\ell}$,
$w^{(d)}_1$, $w^{(d)}_2,\:...,$ parametrizing 
%is equivalent to the algorithm for finding the parameters of 
the effective range function, are found by minimizing the sum of squares of 
deviation of the sets of~$E^{th}_{i}$ 
and~$E_{i}$ with weights enhancing the contribution of energies obtained with 
larger~$N_{\max}$ values,
\begin{equation}
\Xi_{w}=\sqrt{\frac{1}{p}\sum_{i=1}^p{\left(\left(E_{i}^{th}-E_{i}\right)^{2}\left(\frac{N_{\max}^i}{N_M}\right)^2\right)}}.
\label{rms_w}
\end{equation} 
Here $ p$ is the number of energy values and $N_M$ is  the largest value of $N_{\max}^i$ used
in the fit. With the optimal set of the fit parameters $w^{(n)}_1$, $w^{(n)}_2,\:...\,$, $a_{\ell}$,
$w^{(d)}_1$, $w^{(d)}_2,\:...$  we can use Eq.~\eqref{EffRadius}  and~\eqref{PhaseSSHORSE} 
to obtain a parametrization of the~$\hbar\Omega$ dependencies of the eigenenergies~$E_{i}$ in any basis 
space~${\mathbb N}^{i}$.
%More details about this minimization procedure can be found in
%%is described in 
%Ref.~\cite{SSHORSE-K}.

The $S$-matrix and the effective range function~$K(E)$ are related by a simple 
analytic formula.
Therefore, after {obtaining} an accurate parametrization of~$K(E)$, one can search 
numerically
for the $S$-matrix poles in the complex energy plain. Some tricks useful to 
design a stable and fast
numerical algorithm {for} the pole searches at complex energies, are described in 
Ref.~\cite{SSHORSE-K}.
By locating the $S$-matrix poles, we obtain energies~$E_{r}$ and widths~$\Gamma$ 
of
resonances in the many-body nuclear system.

\begin{figure}[b!]
\psfrag{EMeVqq}{$E_{i}$ [MeV]}
\psfrag{hwMeVqq}{$\hbar\Omega$ [MeV]}
\psfrag{deltaqqqqqqqqqq}{$\delta_{1}(E_{i})$ [degrees]}
\centerline{\includegraphics[width=0.47\textwidth]{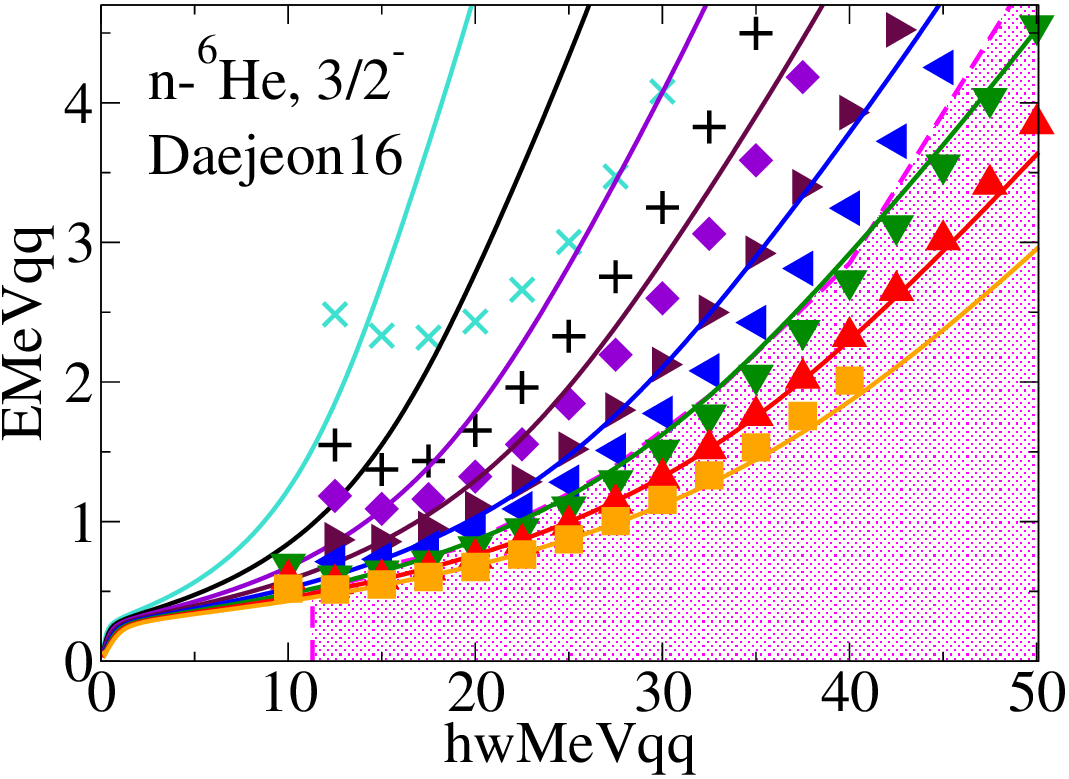}\hfill
\includegraphics[width=0.49\textwidth]{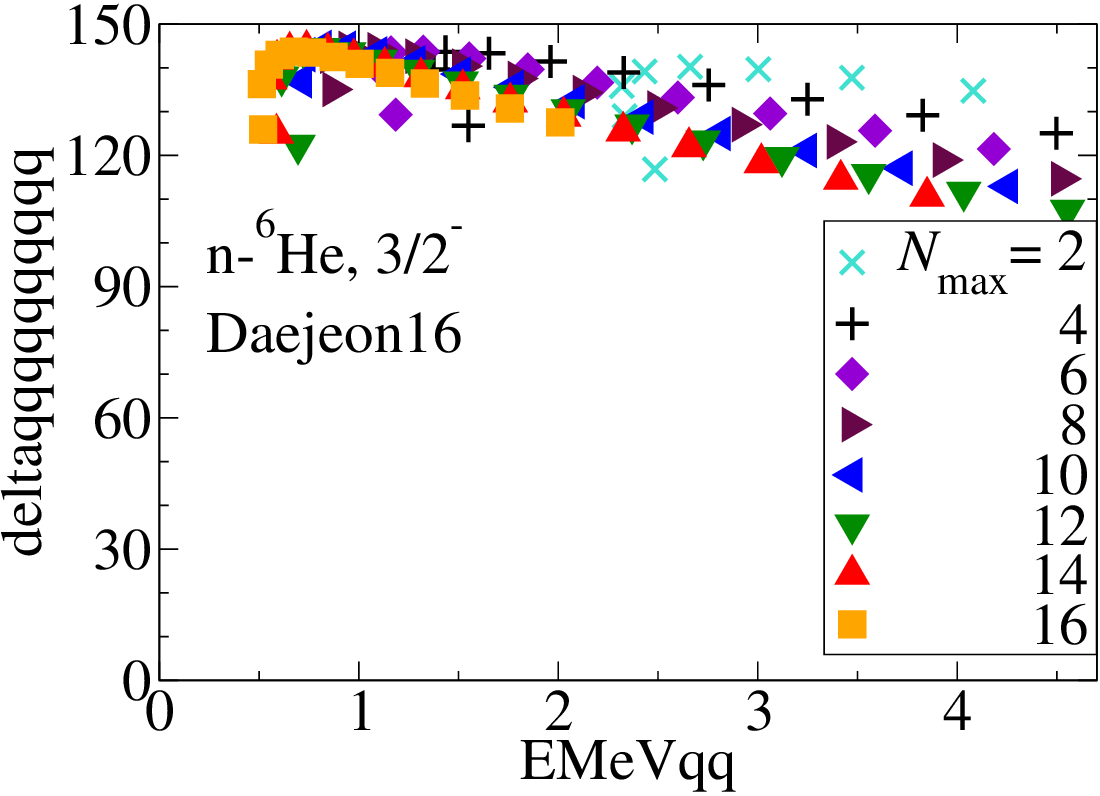}}
\caption{
Left panel: % Points: 
Symbols are  the energies of the relative motion~$E_{i}$ in the %for
 $3/2^-$ scattering 
state obtained in the NCSM with the Daejeon16 $NN$ interaction; the energies used 
for the
SS-HORSE parametrization are taken from the shaded area and the results of the 
SS-HORSE
parametrization of energies for each~$N_{\max}$ are shown by  solid curves of 
respective colors.
%Curves: approximated in SS-HORSE method energies. The pink area involves 
%energies taken in to account in parameterization. 
Right panel:  The phase shifts calculated using Eq.~\eqref{PhaseSSHORSE} at the 
energies
from the left panel.}
%obtained from NCSM energies with the Eq.~\eqref{PhaseSSHORSE}.}
\label{Fig:n-6He32m_Daej}
\vspace{-1mm}      
\end{figure}

\section{\boldmath$n{-}{^6\rm He}$  scattering\label{Sec:n-6He}}

We start from the NCSM calculations of the $^6$He ground state energies $E^6_{i}$ 
with 
the Daejeon16~\cite{Daejeon16} and  JISP16~\cite{JISP16} $NN$ interactions 
with $N_{\max}$ up to 16 and $\hbar\Omega$ ranging from 8 to 50~MeV. Next we 
calculate the lowest eigenenergies $E^7_{i}$ of the $3/2^-$, $1/2^-$, $5/2^-$ and 
$1/2^+$ states in %of 
the $^7$He nucleus with $N_{\max}$ up to 17 with the same 
interactions and the same $\hbar\Omega$ values.

We {first consider} calculations performed with the Daejeon16 $NN$ 
interaction.
%R
The set of the relative motion energies~$E_{i}$ %y
is calculated using Eq.~\eqref{EnergyRelative}.
% with respect formula
%\begin{equation}
%	\label{EnergyRelative2}
%	E(N_{\max},\hbar\Omega)=E^7(N_{\max},\hbar\Omega)-
%	                        E^6(N^\prime_{\max},\hbar\Omega),
%\end{equation}
%where $N^\prime_{\max}=N_{\max}$ for odd parity and $N^\prime_{\max}=N_{\max}-1$ 
%for even parity. Value 
%of $\mathbb N$ defined as
%\begin{equation}
%	{\mathbb N}=N_{\max}+1.
%\end{equation}
As an example, we present in the left
%Left 
panel of Fig.~\ref{Fig:n-6He32m_Daej} %presents by color points NCSM 
the set of relative motion energies~$E_{i}$ in the~$3/2^-$ state.
% obtained with the Daejeon16 $NN$ interaction.
%energies of relative motion. R
The right panel of %this
the same figure presents the set of the phase shifts~$\delta_{\ell}(E_{i})$
at these energies calculated using Eq.~\eqref{PhaseSSHORSE}.
% the same energies from left panel. P

As stated in 
Refs.~\cite{SSHORSE, SSHORSE_PEPAN,Blkh,Blokh, SSHORSE-K,tetran}, we
cannot use all energies~$E_{i}$ obtained by the NCSM for the further SS-HORSE 
analysis.
The set of acceptable energies~$E_{i}$ should be selected for the SS-HORSE.
In particular, the SS-HORSE equations are consistent only with those energies 
obtained at any
given~$N_{\max}$ which increase with~$\hbar\Omega$, i.\:e., for any 
given~$N_{\max}$ we
should have~$\frac{dE}{d\hbar\Omega}>0$. In other words, from the set of 
energies~$E^{N_{\max}}_{i}$ 
obtained by NCSM
with any~$N_{\max}$ we should select only those which are obtained 
with~${\hbar\Omega>\hbar\Omega^{N_{\max}}_{\min}}$, 
where~${\hbar\Omega^{N_{\max}}_{\min}}$ 
corresponds to the minimum of the~$\hbar\Omega$ dependence of the relative motion
energies~$E^{N_{\max}\!}_{i}$.

Next, for the effective range function parametrization, we should select only the 
results obtained
with large enough~$N_{\max}$ and in the ranges of~$\hbar\Omega$ values 
for each~$N_{\max}$ where the phase shifts
converge, at least, approximately. The phase shift convergence means that
the phase shifts~$\delta_{\ell}(E_{i})$ obtained with 
different~$N_{\max}$ and~$\hbar\Omega$
values form a single smooth curve as a function of energy. In the right panel of 
Fig.~\ref{Fig:n-6He32m_Daej}, {we see} that the phase 
shifts~$\delta_{\ell}(E_{i})$
%The phase shifts are seen to 
tend to form a
%the 
smooth curve %with increasing 
as~$N_{\max}$ increases in a range of moderate energies which correspond to
moderate~$\hbar\Omega$ values. The 
phase shifts~$\delta_{\ell}(E_{i})$ obtained with small enough~$N_{\max}$ deviate 
{significantly} from this
single curve in large energy {intervals.  Correspondingly, } the phase shifts obtained even with large~$N_{\max}$ at small energies
 corresponding to small~$\hbar\Omega$ values before
the minima of the $\hbar\Omega$ dependences of~$E^{N_{\max}}_{i}$ also  deviate from the
phase shift curve formed by the NCSM results from other~$N_{\max}$ values. 
%Some deviation 
%from
%a single smooth phase shift curve is seen also for~$\delta_{\ell}(E_{i})$ 
%obtained with the
%largest available~$N_{\max}$ at large enough energies corresponding to the largest
%available~$\hbar\Omega$ values.

\begin{figure}[t]
\psfrag{EMeVqqq}{$E$ [MeV]}
\psfrag{deltadegreesqqq}{$\delta_{\ell}(E)$ [degrees]}
%\parbox[b]{.49\textwidth}{{\includegraphics[width=0.59\textwidth]{n-6He32m12m52m12p_Daejeon16_v1_deltaE.eps}}}
\parbox[b]{.49\textwidth}{{\includegraphics[width=0.59\textwidth]{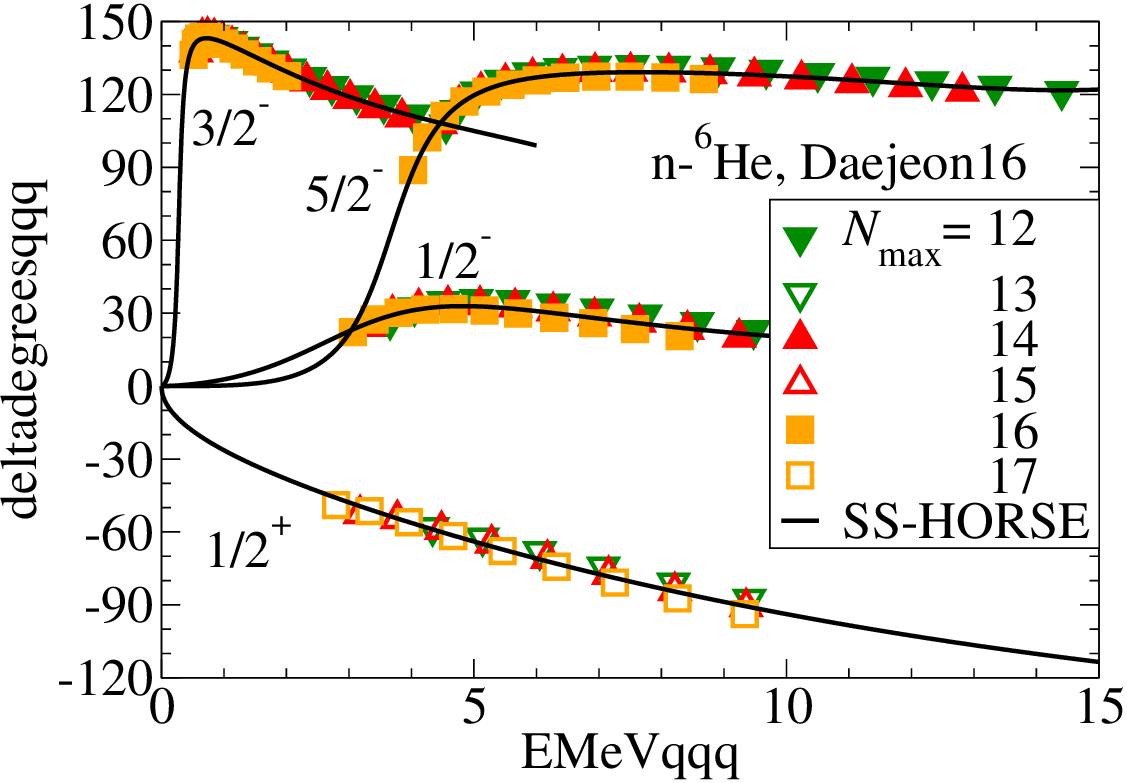}}}
\hfill\parbox[b]{.36\textwidth}{\caption{The phase shifts in the %obtained for the
$3/2^-$, $1/2^-$, $5/2^-$ and $1/2^+$ scattering states obtained with the 
Daejeon16 
$NN$ interaction. Symbols are the selected phase shifts~$\delta_{\ell}(E_{i})${;} 
the SS-HORSE fit
of the phase shifts is 
%Points denotes phase shifts calculated from NCSM 
%energies through~\eqref{PhaseSSHORSE}. The approximated phase shifts are 
presented by black curves.}
\label{Fig:n-6He_Daej}}
%\vspace{-5mm} 
\end{figure}

The energies selected for the SS-HORSE fit are shown by the shaded area in the 
left panel of 
Fig.~\ref{Fig:n-6He32m_Daej}. The solid curves in this panel show the 
parametrization of the NCSM
energies through the function~\eqref{Pade} with a set of fitted parameters. The 
selected energies
produce a set of the phase shifts~$\delta_{1}(E_{i})$ forming a smooth single 
curve, as is seen in
Fig.~\ref{Fig:n-6He_Daej}, where we also present the SS-HORSE $3/2^{-}$ phase 
shifts {accurately} describing 
the set of the selected phase shifts~$\delta_{1}(E_{i})$.

%lying on this curve in phase shifts plot are located
%inside pink area in left panel of Fig.~\ref{Fig:n-6He32m_Daej}, while the  
%corresponding to phase shifts are plotted on 
%Fig.~\ref{Fig:n-6He_Daej}. Energies in pink area were 
%approximated and we present obtained energies in left side of 
%Fig.~\ref{Fig:n-6He32m_Daej} by curves. Parameterization process explained 
%in our previous papers, see e.~g.~Ref.~\cite{SSHORSE-K}. This 
%approximated energies provide phase shifts that are plotted by black 
%curve in Fig.~\ref{Fig:n-6He_Daej}.

We note that we {perform a few} alternative selections of 
energies~$E_{i}$, e.\:g., we
exclude from the selection {some large} energies~$E_{i}$ which lie far
from the resonance. These
alternative energy selections are used for estimating uncertainties of our 
predictions for the  parameters of
the resonance and low-energy scattering.
The resonance energies~$E_{r}$ (relative to the $n+{^6\rm He}$ threshold) 
and widths~$\Gamma$ of resonances in the $^{7}$He nucleus
obtained by a numerical location of the $S$-matrix poles are presented in 
Table~\ref{Tab:resonance} as well as
%together with 
the low-energy scattering parameters, the scattering 
length~$a_\ell$ and the effective
range~$r_\ell$, together with their estimated uncertainties. 
For comparison, we present in Table~\ref{Tab:resonance} also the resonance 
parameters
from the GSM %Gamow shell model 
studies of Ref.~\cite{GSM} {and %from 
the NCSMC studies of 
Refs.~\cite{Navratil7He1, Navratil7He2} with SRG-evolved N$^3$LO chiral $NN$ forces together
with available} %and from  the compilation of the 
experimental data. %~\cite{Cao}. 
Our results for the 
$3/2^-$ resonance are seen to be consistent with the GSM results and 
experiment.

%In the Table~\ref{Tab:resonance} we present resonance energies, widths, 
%scattering lengths and effective radii obtained in NCSM-SS-HORSE for 
%various scattering states. Results obtained in the Gamow shell model 
%(GSM)~\cite{GSM} and results extracted from 
%experimental data~\cite{Tilley} are given for comparison. 
%Our results are close to those from GSM and experiment.

\begin{table}[!t]
\caption{Energies $E_r$ (relative to the $n+{^6\rm He}$ threshold) and widths of  
negative parity % The
 resonant states in $^{7}$He nucleus 
 %energies $E_r$ (counted from $n{-}^6\rm He}$  channel 
%threshold), widths~$\Gamma$, 
and parameters of low-energy scattering $n{-}{^6\rm He}$ in positive and negative 
parity states,
scattering lengths~%s 
$a_\ell$ and effective %radius~%i 
ranges~$r_\ell$, obtained % in the SS-HORSE 
with %the 
Daejeon16 and %the 
JISP16 $NN$ interactions. 
%$D$ is the 
%number of selected NCSM energies used in the fit of effective range function %obtaining resonant and scattering 
%parameters and~$\Xi$ is the rms deviation of the selected energies from the fit. 
Our estimate of the uncertainties of the quoted results are in presented 
parentheses.  
The available results 
of the GSM calculations~\cite{GSM} {and of the NCSMC 
calculations~\cite{Navratil7He1, Navratil7He2} with SRG-evolved N$^3$LO chiral $NN$ force together with }% and 
%Also values calculated in the GSM~\cite{GSM} and in analysis
%compilation of 
experimental 
data are presented for comparison.%
% Note, the resonant state $1/2^-$ have no mentioned in Ref.~\cite{GSM}.
}\vspace{-2.4ex}
\begin{center}
\begin{tabular}{cccccccc}%{cc|c|c|c|c}
\hline
              & Daejeon16 & JISP16    & GSM     & \hspace{-3mm}{NCSMC}& \multicolumn{3}{c}{Experiment}                                    \\
\hline
$3/2^-$       &           &           &         &                          & {\cite{Cao} }  &                       &                      \\
$E_r$, MeV    & 0.27(1)   & 0.70(2)   & 0.39    & \hspace{-2mm}{0.71} & {0.430(3)}    &                       &                      \\
$\Gamma$, MeV & 0.12(1)   & 0.60(2)   & 0.178   & \hspace{-2mm}{0.30} & {0.182(5)}    &                       &                      \\
$a_1$, fm$^3$ & $-$170(10)& $-$66(2)  &         &                          &                     &                       &                      \\
$r_1$, fm$^{-1}$&$-$1.10(3)&$-$0.88(1)&         &                          &                     &                       &                      \\
\hline
$1/2^-$       &           &           &         &                          &{\cite{Wuosmaa}}&{\cite{Boutachkov}}&{\cite{Meister}}\\
$E_r$, MeV    & 2.7(1)    & 2.8(1)    &         & \hspace{-2mm}{2.39} & {3.03(10)}    & {3.53}            & {1.0(1)}       \\
$\Gamma$, MeV & 4.2(1)    & 5.02(2)   &         & \hspace{-2mm}{2.89} & {2}           & {10}              & {0.75(8)}      \\
$a_1$, fm$^3$ & $-$4.0(1) & $-$4.5(2) &         &                          &                     &                        &                     \\
$r_1$, fm$^{-1}$&$-$4.4(2)& $-$3.1(1) &         &                          &                     &                        &                     \\
\hline
$5/2^-$       &           &           &         &                          & {\cite{Tilley}}&                        &                     \\
$E_r$, MeV    & 3.65(2)   & 4.37(4)   & 3.47(2) & \hspace{-2mm}{3.13} & {3.35(10)}    &                        &                     \\
$\Gamma$, MeV & 1.37(1)   & 1.55(2)   & 2.25(28)& \hspace{-2mm}{1.07} & {1.99(17)}    &                        &                     \\
%$a_3$, fm$^5$ & $-$274(4) & $-$119(3) &   
$a_3$, fm$^{7}$  & $-$274(4)    & $-$119(4)   &       &                      &                     &                  &               \\
%$r_3$, fm$^{-3}$&$-$1.22(4)&$-$4.0(1) &     
$r_3$, fm$^{-5}$ & $-$0.0122(4) & $-$0.040(1) &    &                     &                     &                &                \\
\hline
$1/2^+$ \\
$a_0$, fm     & 2.1(2)    & 3.2(5)    &         &                          &                     &                        &                     \\
$r_0$, fm     & 2.1(2)    & 1.1(6)    &         &                          &                     &                        &                     \\
\hline          
\end{tabular}
\end{center}
\label{Tab:resonance}
\vspace{-2.5ex}
\end{table}

The same approach is used to examine the $1/2^{-}$ and $5/2^{-}$ resonances in 
the $^{7}$He
nucleus. The results for the phase shifts together with selected  phase 
shifts~$\delta_{1}(E_{i})$
are also shown in Fig.~\ref{Fig:n-6He_Daej} while the resonance and low-energy 
scattering parameters
are presented in Table~\ref{Tab:resonance}.

%The situation with the $1/2^-$ state appears to be more complicated. 
We note that the %The 
convergence of the~$1/2^-$ phase shits, where we obtain a wide resonance, is 
slower than in the
case of the~% in this case is slow as compared to 
$3/2^-$ state. As a result, our predictions for the~$1/2^-$ resonance energy and 
width
{tend to have larger} uncertainties. {The} predictions for the low-energy 
scattering parameters
{for the $1/2^-$} case appear to
{have uncertainties comparable to the resonance parameter uncertainties.}
%There are smaller number of points forming a smooth 
%curve. The NCSM energies obtained with $12\le N_{\max}\le16$ and 
%$15\le\hbar\Omega\le40$~MeV (excluding the energy obtained with 
%$N_{\max}=14$ and $\hbar\Omega=40$~MeV) we use for the SS-HORSE analysis. 

{The experimental situation for the $1/2^-$ resonance is not clear. While 
the resonant energies of Refs.~\cite{Wuosmaa, Boutachkov} are comparable, the 
widths are very different. Our results are in fair agreement with the NCSMC results and the neutron pickup
and proton-removal reaction experiments~\cite{Wuosmaa} and definitely do not support the 
interpretation of experimental data on one-neutron knockout from $^{8}$He of Ref.~\cite{Meister}
%hypothesis of  
advocating a low-lying
($E_r\sim1$~MeV) narrow ($\Gamma\le1$~MeV) $1/2^-$ resonance in $^{7}$He.}

%As is seen from %the 
%Table~\ref{Tab:resonance}, our results 
%overestimate both the energy and the width of the $1/2^-$ resonance. 
%We plot the calculated phase shifts in Fig.~\ref{Fig:n-6He_Daej}.

In the case of the $5/2^-$ scattering, %state 
the phase 
shifts convergence is similar to that of the~$3/2^-$ state. 
%The NCSM 
%energies obtained in bases with $12\le N_{\max}\le16$ and 
%$15\le\hbar\Omega\le50$~MeV are used in the parameterization. The phase 
%shifts obtained in this case are plotted in Fig.~\ref{Fig:n-6He_Daej}. 
The resonance energy and width %are 
presented in Table~\ref{Tab:resonance} are seen to be reasonably %. They are 
close to the experimental data, GSM {and NCSMC} results.
%{\clr{}We note that in this case the low-energy variation of the effective range 
%function~$K(E)$ is dominated by the quadratic in energy term in 
%Eq.~\eqref{effrengeexp}, the
%effective range~$r_{3}$ is small and it is hard even to fix its sign in our 
%calculations.}

%\mbox{}
%
%{HERE}
%
%\mbox{}

We analyze also the scattering in the $1/2^+$ state in our NCSM-SS-HORSE 
approach. The
$1/2^+$ scattering  phase shifts shown in Fig.~\ref{Fig:n-6He_Daej} monotonically 
decrease 
without any signal of a resonant state.
%Also the $1/2^+$ scattering state is analyzed in the NCSM-SS-HORSE method. 
%In this case, convergence of phase shifts is like $1/2^-$ state therefore 
%we used not so much points in the analysis. All used NCSM results belong to 
%$13\le N_{\max}\le17$ and $15\le\hbar\Omega\le40$~MeV. Nevertheless we 
%obtain phase shifts (see Fig.~\ref{Fig:n-6He_Daej}) that 
%not support any $S$-matrix pole in energy interval under consideration. 
This result is in an agreement with the experimental data and the GSM
predictions of Ref.~\cite{GSM} {and NCSMC 
predictions~\cite{Navratil7He1, Navratil7He2}}. % in GSM.

\begin{figure}[t!]
\psfrag{EMeVqqq}{$E$ [MeV]}
\psfrag{deltadegreesqqq}{$\delta_{\ell}(E)$ [degrees]}
\parbox[b]{.49\textwidth}{{\includegraphics[width=0.59\textwidth]{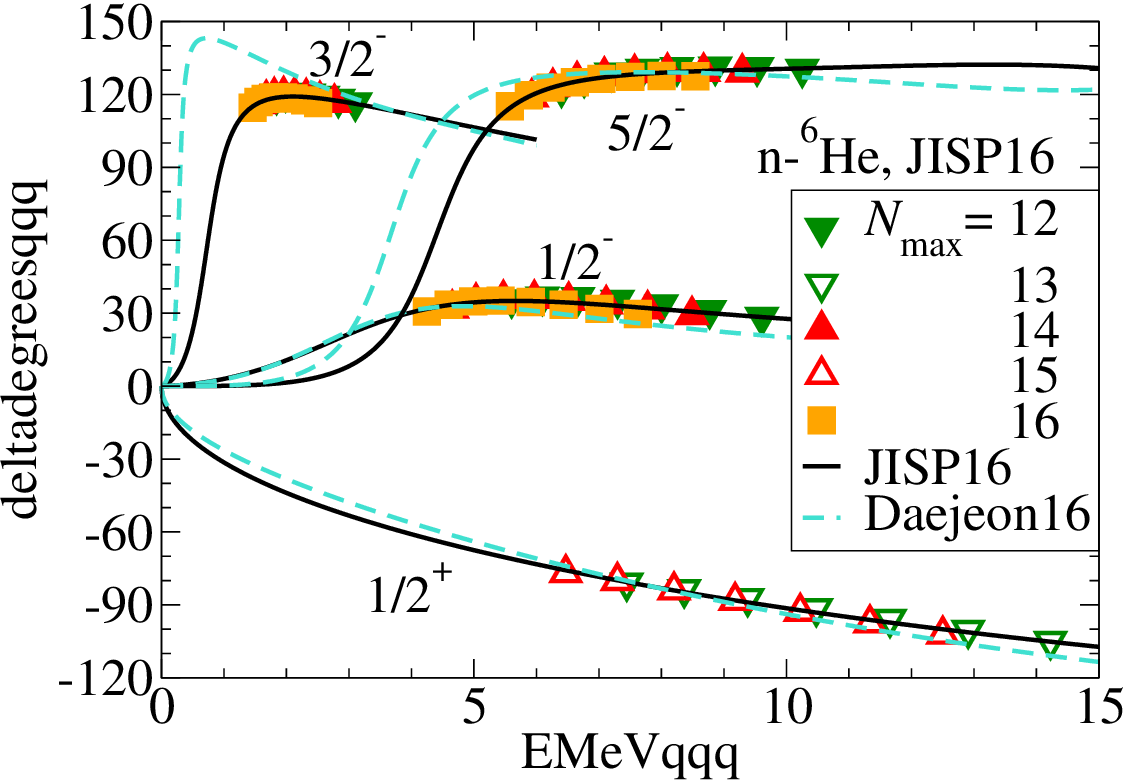}}}
\hfill\parbox[b]{.36\textwidth}{\caption{The phase shifts in the %obtained for 
$3/2^-$, $1/2^-$, $5/2^-$ and $1/2^+$ scattering states obtained 
%in the SS-HORSE method 
with the JISP16 $NN$ interaction in comparison with those obtained with the 
Daejeon16 
(red dashed curves). See
%Denotations are the same with 
Fig.~\ref{Fig:n-6He_Daej} for other details.
%. Obtained with the Daejeon16 phase shifts are 
%presented by red dashed curves.
}
\label{Fig:n-6He_Jisp}}
\vspace{-1mm}
\end{figure}

The phase shifts obtained with the JISP16 $NN$ interaction are compared with 
those from Daejeon16
in Fig.~\ref{Fig:n-6He_Jisp}. The only difference in getting these JISP16 results 
is that we 
avoided the expensive~$N_{\max}=17$ calculations for the positive-parity states
since there is no experimental evidence for the positive-parity resonances in $^{7}$He
and we do not see any indication of such resonances in {our phase shift} calculations. 
The JISP16
and Daejeon16 $1/2^+$ scattering   phase shifts are seen to be very close as
{are} the
respective low-energy scattering parameters listed in Table~\ref{Tab:resonance}. 
The~$3/2^{-}$
and~$5/2^{-}$ $^{7}$He resonances are generated  by the JISP16 {at slightly} higher energies; 
the~$1/2^{-}$ resonance appears approximately at the same energy, however its 
width {is somewhat} larger {in the JISP16 results compared with the Daejeon16 results.}

%The scattering with the the JISP16 $NN$ interaction was considered in the 
%similar manner. In this case we use the NCSM results only up to $N_{\max}=16$ 
%in assumption that results with $N_{\max}=17$ will not affect 
%significantly. Note, that in case of the the JISP16 interaction in analysis 
%are used the NCSM results with $25\le\hbar\Omega\le40$~MeV and all NCSM 
%energies more higher compared with the Daejeon16. Obtained phase shifts are 
%plotted in Fig.~\ref{Fig:n-6He_Jisp} and resonance energies, widths, scattering 
%lengths and effective radii are presented in the Table~\ref{Tab:resonance}.

\section{Summary and conclusions}

%The phase shifts with both $NN$ interactions used in this work are plotted 
%in Fig.~\ref{Fig:n-6He_Jisp}. The results for $1/2^-$ and $1/2^+$ 
%states look the similar with both interaction meanwhile there is 
%significant difference in $3/2^-$ and $5/2^-$ states. First of all 
%the Daejeon16 provides resonances with lower energy. These energies are more 
%close with experimental data. Also in $3/2^-$ the Daejeon16 provide narrower 
%resonance that again more similar with experimental data. The resonant 
%parameters obtained in $1/2^-$ are in a contradiction with experimental 
%data.

We performed a study of the $n+\rm{^{6}He}$ continuum states within the 
single-channel
SS-HORSE extension of the {\it ab initio} NCSM with JISP16 and Daejeon16 $NN$ 
interactions.
No resonance was found in the~$1/2^{+}$ state {consistent} with the 
GSM~\cite{GSM}{, NCSMC~\cite{Navratil7He1, Navratil7He2} studies}
and experimental situation. {The~$1/2^{-}$ resonance is predicted by both 
interactions to be wide enough and at the energy
 in a reasonable agreement with the NCSMC~\cite{Navratil7He1, Navratil7He2} %results.
calculations and results of experiments of  Refs.~\cite{Wuosmaa, Boutachkov}
and clearly contradicts with the hypothesis of a low-lying narrow resonant state suggested in Ref.~\cite{Meister}.
%But experimental background for this resonance is not clear
} 
We note however that this as well as other $^{7}$He resonances are known from the 
experiment with weak spin-parity assignment arguments.
Our results for the narrow~$3/2^{-}$ and
wide~$5/2^{-}$ resonances are in a reasonable agreement with experiment {and with} 
results
quoted in the GSM~\cite{GSM} {and NCSMC~\cite{Navratil7He1, Navratil7He2} studies}{. However,} JISP16 overestimates the width of 
the~$3/2^{-}$ and
the energy of the~$5/2^{-}$ resonances. 

\section*{Acknowledgements}

This work is supported in part  by the National Research Foundation of 
Korea (NRF) grant funded by the Korea government (MSIT)
(No.~2018R1A5A1025563), 
by the %\linebreak 
Russian Science Foundation under Grant No.~16-12-10048, 
by the U.S.\ Department of Energy under Grants
No.~DESC00018223 (SciDAC/NUCLEI) and %\linebreak 
No.~%\mbox
{DE-FG02-87ER40371}, 
%by the US National Science Foundation under Grant No. 1516181, 
by the Rare Isotope Science Project of the Institute for Basic Science funded by 
Ministry of Science and ICT and National Research Foundation of Korea 
(2013M7A1A1075764). 
Computational resources were provided by the National Energy Research 
Scientific Computing Center (NERSC), which is supported by the Office 
of Science of the U.S.\ Department of Energy under Contract 
No.~DE-AC02-05CH11231, and by the 
National Supercomputing Center of Korea with supercomputing resources including technical  
support (KSC-2018-COL-0002).
%Supercomputing Center/Korea Institute of 
%Science and Technology Information including technical
%support (KSC-2015-C3-003).

\end{document}